\documentstyle[12pt]{article}
\begin{document}
\baselineskip 18pt

\def\today{\ifcase\month\or
 January\or February\or March\or April\or May\or June\or
 July\or August\or September\or October\or November\or December\fi
 \space\number\day, \number\year}

\def\thebibliography#1{\section*{References\markboth
 {References}{References}}\list
 {[\arabic{enumi}]}{\settowidth\labelwidth{[#1]}
 \leftmargin\labelwidth
 \advance\leftmargin\labelsep
 \usecounter{enumi}}
 \def\newblock{\hskip .11em plus .33em minus .07em}
 \sloppy
 \sfcode`\.=1000\relax}
\let\endthebibliography=\endlist
\def\lsim{\ ^<\llap{$_\sim$}\ }
\def\gsim{\ ^>\llap{$_\sim$}\ }
\def\r2{\sqrt 2}
\def\beq{\begin{equation}}
\def\eeq{\end{equation}}
\def\beqn{\begin{eqnarray}}
\def\eeqn{\end{eqnarray}}
\def\rmuu{\gamma^{\mu}}
\def\rmud{\gamma_{\mu}}
\def\PL{{1-\gamma_5\over 2}}
\def\PR{{1+\gamma_5\over 2}}
\def\sinW2{\sin^2\theta_W}
\def\AEM{\alpha_{EM}}
\def\mul{M_{\tilde{u} L}^2}
\def\mur{M_{\tilde{u} R}^2}
\def\mdl{M_{\tilde{d} L}^2}
\def\mdr{M_{\tilde{d} R}^2}
\def\mz2{M_{z}^2}
\def\c2b{\cos 2\beta}
\def\au{A_u}
\def\ad{A_d}
\def\cob{\cot \beta}
\def\v#1{v_#1}
\def\tb{\tan\beta}
\def\epem{$e^+e^-$}
\def\KK{$K^0$-$\bar{K^0}$}
\def\wi{\omega_i}
\def\xj{\chi_j}
\def\Wmu{W_\mu}
\def\Wnu{W_\nu}
\def\m#1{{\tilde m}_#1}
\def\mH{m_H}
\def\mw#1{{\tilde m}_{\omega #1}}
\def\mx#1{{\tilde m}_{\chi^{0}_#1}}
\def\mc#1{{\tilde m}_{\chi^{+}_#1}}
\def\mwi{{\tilde m}_{\omega i}}
\def\mxi{{\tilde m}_{\chi^{0}_i}}
\def\mci{{\tilde m}_{\chi^{+}_i}}
\def\mz{M_z}
\def\sw{\sin\theta_W}
\def\cw{\cos\theta_W}
\def\cb{\cos\beta}
\def\sb{\sin\beta}
\def\rwi{r_{\omega i}}
\def\rxj{r_{\chi j}}
\def\rfp{r_f'}
\def\Kik{K_{ik}}
\def\Fq2{F_{2}(q^2)}
\begin{titlepage}

\  \
\vskip 0.5 true cm 
\begin{center}
{\large {\bf The Neutron and the Lepton EDMs in MSSM, }}\\  
{\large {\bf Large CP violating Phases, and the Cancellation Mechanism }}  
\vskip 0.5 true cm
\vspace{2cm}
\renewcommand{\thefootnote}
{\fnsymbol{footnote}}
 Tarek Ibrahim$^{a,b}$ and Pran Nath$^a$  \\

\end{center}
\vskip 0.5 true cm 
\noindent
{a. Department of Physics, Northeastern University}  \\
{Boston, MA 02115, USA } \\
{b. Permanent address: Department of  Physics, Faculty of Science}\\
{University of Alexandria, Alexandria, Egypt}\\ 


\vskip 1.0 true cm

\centerline{\bf Abstract}
\medskip
An analysis of the electric dipole moment (EDM) of the neutron  and
of the leptons 
in the minimal supersymmetric standard model (MSSM) with
the most general allowed set of CP violating phases without 
generational mixing is given. 
The analysis includes the contributions from the gluino, the chargino
and the neutralino exchanges to the electric dipole operator,  the  
chromoelectic dipole operator, and the CP violating purely gluonic 
dimension six operator.
It is found that the EDMs depend only on certain combination of the CP 
phases. The independent set of such phases is classified. 
The analysis of the EDMs given here provides the 
framework for the exploration of the effects of large 
 CP violating phases on low energy phenomena such as the search
 for supersymmetry at colliders, 
  and in the analyses of dark matter consistent with the 
  experimental limits on EDMs via the mechanism of internal cancellations.
 
\end{titlepage}

\newpage 
It is well known that supersymmetric theories contain many new 
sources of CP violation and can produce large contributions to the
electric dipole moments of the neutron and of the electron[1-5].
With normal size CP violating phases, i.e., 
phases O(1), and with SUSY spectrum in the TeV range, the neutron 
and the electron EDMs already lie in excess of the current experimental 
limit,
which for the neutron is\cite{altra} 
$d_n < 1.1\times 10^{-25}$ ecm
and for the electron is \cite{commins}
$d_e$ $<$ $4.3\times 10^{-27}$ ecm.
Two approaches have usually been adopted to rectify this  
situation. The first is to make the phases small, i.e.
O($10^{-2}$)\cite{ellis}, 
and the other is to use mass suppression by making 
the SUSY spectrum heavy, i.e., in the several TeV range\cite{na}. The
first case, however, represents fine tuning, while the second  violates 
naturalness and also makes
the SUSY spectrum so heavy that it may not be accesssible even
at the LHC. Recently, a third possibility was proposed\cite{us1},
 i.e., that   
of internal cancellations in EDMs reducing them below the experimental
limits even for CP violating phases O(1). 

In recent works the importance of CP violating phases on low
	energy phenomenona has been recognized\cite{falk1,kane,falk2}.
	In ref.\cite{kane} it is shown that large CP violating phases can affect
	sparticle searches at colliders and in 
	 ref.\cite{falk2} it is found that large CP 
	violating phases can produce large effects on the neutralino relic
	density consistent with the experimental constraints on the 
	the neutron and on the electron EDM via the cancellation
	mechanism\cite{us1}. However, as one goes beyond the
	framework of minimal supergravity to include non-universalities
	in the soft SUSY breaking parameters one finds that new 
	CP violating phases arise which also affect low energy phenomena.
	Currently the effect of CP violating phases beyond the two CP
	phases allowed by the minimal supergravity cannot be 
	investigated because the analytic computations of the EDMs in 
	terms of these phases do not exist in the literature 
	and consequently the EDM constraints arising from experiment  
	cannot be implemented. The purpose of this Letter is to  provide 
	the analytic results for the EDMs beyond the minimal supergravity
	model by inclusion of all CP  
 	violating phases in the framework of MSSM. As is conventional we ignore generational
 	mixings whose effects are known to be small. 
 	 We analyse all one loop diagrams 
 	 with the gluino, the chargino, and the neutralino exhanges for the
 	 electric dipole, and the chromoelectric dipole operators
 	 allowing for all phases. We also analyse the two 
 	 loop diagrams which contribute to the  purely gluonic  
	dimension six operator allowing again for all CP 
	violating phases. 

In MSSM the CP violating phases relevant for the analysis of the EDM's 
arise from the soft SUSY breaking sector of the theory. We display this
sector below\cite{applied}:
\beqn
V_{SB} & &={m_1^2|H_1|^2+m_2^2|H_2|^2 -
  [B \mu\epsilon_{ij} H_1^i H_2^j+H.c.]} \nonumber\\
& &
\hspace{4cm} +{M_{\tilde{Q}}^2[\tilde{u}_{L}^*\tilde{u}_{L}+
\tilde{d}_{L}^*\tilde{d}_{L}]+M_{\tilde{U}}^2 \tilde{u}_{R}^*\tilde{u}_{R}+
M_{\tilde{D}}^2 \tilde{d}_{R}^*\tilde{d}_{R}}\nonumber\\
& &
{+M_{\tilde{L}}^2[\tilde{\nu}_{e}^*\tilde{\nu}_{e}+
\tilde{e}_{L}^*\tilde{e}_{L}]
+M_{\tilde{E}}^2 \tilde{e}_{R}^*\tilde{e}_{R}}\nonumber\\
& &
{+\frac{g m_0}{\r2 m_W} \epsilon_{ij}[\frac{m_e A_e}{\cb} H_1^i \tilde{l}_{L}^j
\tilde{e}_{R}^* +\frac{m_d A_d}{\cb} H_1^i \tilde{q}_{L}^j
\tilde{d}_{R}^* -\frac{m_u A_u}{\sb} H_2^i \tilde{q}_{L}^j
\tilde{u}_{R}^*+H.c. ]}\nonumber\\
& &
{+\frac{1}{2}[\tilde{m}_3  \bar{\tilde{g}}e^{-i\gamma_5\xi_3} \tilde{g} 
   +\m2 \bar{\tilde{W}}^a e^{-i\gamma_5\xi_2}
   \tilde{W}^a+\m1 \bar{\tilde{B}}  e^{-i\gamma_5\xi_1}  \tilde{B}]
   +\Delta V_{SB}}
\eeqn
where  ($\tilde{l}_{L}$, $\tilde{q}_{L}$) are the SU(2) (slepton, squark) 
doublets,  tan$\beta$=$|<H_2>/<H_1>|$ where $H_2$ gives mass to the 
up quark and $H_1$ gives mass to the down quark and the lepton, 
 $\Delta V_{SB}$ is
the one loop contribution to the effective potential,
and we have suppressed the generation indices. In the above, 
$A_{u}$, $A_d$, $A_e$, $\mu$ and B are all complex.
 Additionally,  after spontaneous breaking
of the electro-weak symmetry the vacuum expectation values of the  
Higgs fields are in general complex. Some of the phases in Eq.(1) 
can be eliminated by field redefinitions. However, the choice of which
ones to eliminate is arbitrary. Rather, in our analysis we carry all the
phases to the end and our final expressions 
contain only certain specific combinations.

	One defines the EDM of a spin-$\frac{1}{2}$ particle by  the 
	effective lagrangian
	\beq
{\cal L}_I=-\frac{i}{2} d_f \bar{\psi} \sigma_{\mu\nu} \gamma_5 \psi F^{\mu\nu}
\eeq
In the following we will compute the contributions of the gluino,
the chargino and the neutralino exchanges in MSSM
 keeping all CP violating phases.

The gluino sector contains a phase $\xi_3$ in the gluino mass term. We 
make a transformation on the gluino field to move this phase from the 
mass term to the 
quark-squark-gluino vertex which  
    is then given by ~\cite{applied}:
\beq
-{\cal L}_{q-\tilde{q}-\tilde{g}}=\r2 g_s T_{jk}^a \sum_{i=u,d} 
        (-e^{-i\xi_3/2}\bar{q}_{i}^j \PL \tilde{g}_a \tilde{q}_{iR}^k +
	e^{i\xi_3/2}\bar{q}_{i}^j \PR \tilde{g}_a \tilde{q}_{iL}^k) + H.c. ,
\eeq
Here $j,k=1-3$ are the quark and the
squark color indices, $a=1-8$ are the gluino color indices,  and 
$T_a$ are the SU(3)$_C$ generators.  
The scalar fields $\tilde{q}_L$ and $\tilde{q}_R$    
are in general linear combinations of the mass eigenstates $\tilde q_{i}$
(i=1,2) so that 
\beq
\tilde{q}_L=D_{q11} \tilde{q}_1 +D_{q12} \tilde{q}_2,~~
\tilde{q}_R=D_{q21} \tilde{q}_1 +D_{q22} \tilde{q}_2
\eeq
where $D_{qij}$ are the matrices that diagonalize the squark matrix
such that 
$D_{q}^\dagger$ $M_{\tilde{q}}^2$ $D_q$=${\rm diag}(M_{\tilde{q}1}^2,
              M_{\tilde{q}2}^2)$, 
where
\beq
M_{\tilde{q}}^2=\left(\matrix{M_{\tilde{Q}}^2+m{_q}^2+M_{z}^2(\frac{1}{2}-Q_q
\sin^2\theta_W)\cos2\beta & m_q(A_{q}^{*}m_0-\mu R_q) \cr
   	          	m_q(A_{q} m_0-\mu^{*} R_q) & M_{\tilde{U}
}^2+m{_q}^2+M_{z}^2 Q_q \sin^2\theta_W \cos2\beta}
		\right)
\eeq
Here $Q_u=2/3(-1/3)$ for q=u(d), $R_q=v_1/v_2^* (v_2/v_1^*)$ for q=u(d), 
 and one parametrizes $D_q$ so that 
\beq
D_q=\left(\matrix{\cos \frac{\theta_q}{2} 
           & -\sin \frac{\theta_q}{2} e^{-i\phi_{q}} \cr
	   \sin \frac{\theta_q}{2} e^{i\phi_{q}}
		&\cos \frac{\theta_q}{2}}
		\right),
\eeq
where $ M_{\tilde{q}21}^2=|M_{\tilde{q}21}^2| e^{i\phi_{q}}$
and we choose the range of $\theta_q$ so that 
${{-\pi}\over {2}} \leq  \theta_q \leq {{\pi}\over {2}}$ where 
$\tan \theta_q=
\frac{2|M_{\tilde{q}21}^2|}{M_{\tilde{q}11}^2-M_{\tilde{q}22}^2}$.  	 
In terms of the mass eigenstates $\tilde{q}_1$ and  $\tilde{q}_2$ 
  the gluino contribution to the EDM of the quark is given by 
\beq
{d_{q-gluino}^E}/{e}=\frac{-2 \alpha_{s}}{3 \pi}  m_{\tilde{g}}Q_{\tilde{q}} 
{\rm Im}(\Gamma_{q}^{11}) [\frac{1}{M_{\tilde{q}1}^2}
 {\rm B}(\frac{m_{\tilde{g}}^2}{M_{\tilde{q}1}^2}) -\frac{1}{M_{\tilde{q}2}^2}
{\rm B}(\frac{m_{\tilde{g}}^2}{M_{\tilde{q}2}^2})].
\eeq
where $\Gamma_{q}^{1k}=e^{-i\xi_3} D_{q2k} D_{q1k}^*$, 
  $\alpha_s$=${g_{s}^2}\over {4\pi}$, $m_{\tilde{g}}$ is the gluino mass,
   and  $B(r)=(2(r-1)^2)^{-1}(1+r+2rlnr(1-r)^{-1})$. An explicit
   analysis gives  $\Gamma_{q}^{12}=-\Gamma_{q}^{11}$ where

\beq
{\rm Im}(\Gamma_{q}^{11})=\frac{m_q}{M_{\tilde{q}1}^2-M_{\tilde{q}2}^2}
        (m_0 |A_q| \sin (\alpha_q -\xi_3)+ |\mu| \sin 
        (\theta_{\mu}+\chi_1+\chi_2+\xi_3) |R_q|),
\eeq
which holds for both signs of
 $M_{\tilde{q}1}^2-M_{\tilde{q}2}^2$, and 
  the phases $\chi_i$ (i=1,2) are defined so that
$v_i=<H_i>=|v_i|e^{i\chi_i}$(i=1,2). From Eq.(8) we see that the
 combinations of phases that enter  are ($\alpha_q$-$\xi_3)$ and
 $\xi_3+\theta_{\mu}+\chi_1+\chi_2$, or alternately one can choose
 them to be $\alpha_q+\theta_{\mu}+\chi_1+\chi_2$ and 
 $\xi_3+\theta_{\mu}+\chi_1+\chi_2$.

To discuss the contribution of the chargino exchanges  we  begin by
exhibiting the  chargino mass matrix  

\beq
M_C=\left(\matrix{|\m2|e^{i\xi_2} & \r2 m_W  \sb e^{-i\chi_2}\cr
	\r2 m_W \cb e^{-i\chi_1}& |\mu| e^{i\theta_{\mu}}}
            \right)
\eeq
It is useful to define the transformation
$ M_C=B_R M_C'B_L^{\dagger}$ 
so that 
\beq
M_C'=\left(\matrix{|\m2| & \r2 m_W  \sb \cr
	\r2 m_W \cb & |\mu| e^{i(\theta_{\mu}+\xi_2+\chi_1+\chi_2)}}
            \right)
\eeq
where $B_R=diag(e^{i\xi_2},e^{-i\chi_1})$ and 
$B_L=diag(1,e^{i(\chi_2+\xi_2)})$.
The matrix $M'_C$  can be diagonalized by the biunitary tranformation 
$U_R^{\dagger}M_C'U_L$=diag
$(\mc1, \mc2)$.
It is clear that the matrix elements of $U_L$ and $U_R$  are functions only
of the combination $\theta=\theta_{\mu}+\xi_2+\chi_1+\chi_2$.
 We also have
$U^* M_C V^{-1}=diag(\mc1,\mc2)$ where
$U=(B_R U_R)^T$, and 
V=$(B_LU_L)^\dagger$. Using the fermion-sfermion-chargino
interaction  we find that 
the chargino contribution to the EDM for the up quark is  as follows 
\beq
{d_{u-chargino}^{E}}/{e}=\frac{-\AEM}{4\pi\sinW2}\sum_{k=1}^{2}\sum_{i=1}^{2}
      {\rm Im}(\Gamma_{uik})
               \frac{\mci}{M_{\tilde{d}k}^2} [Q_{\tilde{d}}
                {\rm B}(\frac{\mci^2}{M_{\tilde{d}k}^2})+
	(Q_u-Q_{\tilde{d}}) {\rm A}(\frac{\mci^2}{M_{\tilde{d}k}^2})],
\eeq
\noindent
where $A(r)=(2(1-r)^{-2}(3-r+2lnr(1-r)^{-1})$ and 
\beq
\Gamma_{uik}=\kappa_u V_{i2}^* D_{d1k} (U_{i1}^* D_{d1k}^*-
		\kappa_d U_{i2}^* D_{d2k}^*)
\eeq   
and
\beqn
\kappa_u=\frac{m_ue^{-i\chi_2}}{\r2 m_W \sb}, 
 ~~\kappa_{d,e}=\frac{m_{d,e}e^{-i\chi_1}}{\r2 m_W \cb}
\eeqn
Substitution of the form of $U$ and $V$ matrices gives:
\beq
\Gamma_{ui1(2)}=|\kappa_u| (cos^2 \theta_d /2) [U_{L2i} U^*_{R1i}]
-(+) \frac{1}{2} |\kappa_u \kappa_d| (sin \theta_d)
[U_{L2i}U^*_{R2i}] e^{i\{\xi_2 -\phi_d\}}
\eeq
The terms between the brackets $[ ~]$ in Eq.(14) are functions of $\theta$ and 
from the
definition of $\theta_d$ (as given in the text following Eq.(6)) 
the terms between the brackets $()$ in Eq.(14) are functions
of the combination $\alpha_d+\theta_{\mu}+\chi_1+\chi_2$. By taking the
imaginary part of $\Gamma$ and using the definition of $\phi_d$ (as given 
in the text following Eq.(6)) one can 
show that ($\xi_2-\phi_d$) depends on the combinations ($\xi_2-\alpha_d$),
($\xi_2+\theta_{\mu}+\chi_1+\chi_2$)
and ($\alpha_d+\theta_{\mu}+\chi_1+\chi_2$). 
So we are left only with the
two combinations $\alpha_d+\theta_{\mu}+\chi_1+\chi_2$ and
$\xi_2+\theta_{\mu}+\chi_1+\chi_2$ with $\xi_2-\alpha_d$ being just a 
linear combination of the first two.
 Similar analyses hold for the chargino
contributions to the down quark and one gets only two phase combinations
which are identical to the case above with $\alpha_d$ replaced by 
$\alpha_u$. For the case of the charged lepton we find 

\beq
{d_{e-chargino}^{E}}/{e}=\frac{\AEM}{4\pi\sinW2} 
    {m_{\tilde{\nu}e}^2} \sum_{i=1}^{2} \mci {\rm Im}
     (\Gamma_{ei})   
	{\rm A}(\frac{\mci^2}{m_{\tilde{\nu}e}^2})
\eeq
where 
$\Gamma_{ei}=
 (\kappa_e U_{i2}^* V_{i1}^*)=|\kappa_e| U_{R2i}^*U_{L1i}$. 
A direct inspection of $\Gamma_{ei}$ shows that it depends on only 
one combination,
i.e., $\xi_2$+$\theta_{\mu}$+$\chi_1$+$\chi_2$. 
 
In order to discuss the neutralino exchange contributions we first exhibit
the neutralino mass matrix $M_{\chi^0}$ with the most general allowed set of 
CP violating phases

\beq
\left(\matrix{|\m1|e^{i\xi_1}
 & 0 & -\mz\sw\cb e^{-i\chi_1} & \mz\sw\sb e^{-i\chi_2} \cr
  0  & |\m2| e^{i\xi_2} & \mz\cw\cb e^{-i \chi_1}& -\mz\cw\sb e^{-i\chi_2} \cr
-\mz\sw\cb e^{-i \chi_1} & \mz\cw\cb e^{-i\chi_2} & 0 &
 -|\mu| e^{i\theta_{\mu}}\cr
\mz\sw\sb e^{-i \chi_1} & -\mz\cw\sb e^{-i \chi_2} 
& -|\mu| e^{i\theta_{\mu}} & 0}
			\right).
\eeq
Next we  make the transformation 
$M_{\chi^0}$=$P_{\chi^0}^T$ $M_{\chi^0}'$ $P_{\chi^0}$
where 
\beq
P_{\chi^0}=diag(e^{i\frac{\xi_1}{2}},e^{i\frac{\xi_2}{2}},e^{-i
(\frac{\xi_1}{2}
+\chi_1)},
e^{-i(\frac{\xi_2}{2}+\chi_2)})
\eeq
After the transformation the matrix $M_{\chi^0}^{'}$ takes the form
\beq
\left(\matrix{|\m1| & 0 & -\mz\sw\cb & \mz\sw\sb
                  e^{-i\frac{\Delta \xi}{2}} \cr
    0  & |\m2| & \mz\cw\cb e^{i\frac{\Delta \xi}{2}} & -\mz\cw\sb \cr
-\mz\sw\cb & \mz\cw\cb e^{i\frac{\Delta \xi}{2}} & 0 & -|\mu|e^{i\theta '} \cr
 \mz\sw\sb e^{-i\frac{\Delta \xi}{2}}  & -\mz\cw\sb & -|\mu|e^{i\theta '} & 0}
			\right).
\eeq
where $\theta^{'} =\frac{\xi_1+\xi_2}{2}+\theta_{\mu}+\chi_1+\chi_2$, 
and  $\Delta \xi=(\xi_1-\xi_2)$.
 Now the matrix $M_{\chi^0}^{'}$
can be diagonalized by the transformation
$Y^T M_{\chi^0}' Y$=${\rm diag}(\mx1, \mx2, \mx3, \mx4)$.
It is clear that the transformation matrix Y is a function only of 
$\theta^{'}$ and $\Delta \xi/2$. Combining our results we find that the 
 complex non hermitian and symmetric matrix $M_{\chi^0}$  
can be  diagonalized  using a unitary matrix $X=P_{\chi^0}^{\dagger}Y$ 
such that
$X^T M_{\chi^0} X$=${\rm diag}(\mx1, \mx2, \mx3, \mx4)$.
We can now write down the
 neutralino exchange contribution to the fermion EDM as follows:
\beq
{d_{f-neutralino}^E}/{e}=\frac{\AEM}{4\pi\sinW2}\sum_{k=1}^{2}\sum_{i=1}^{4}
{\rm Im}(\eta_{fik})
               \frac{\mxi}{M_{\tilde{f}k}^2} Q_{\tilde{f}}
{\rm B}(\frac{\mxi^2}{M_{\tilde{f}k}^2})
\eeq 
where
\beqn
\eta_{fik} & &={(a_0 X_{1i} D_{f1k}^*
  + b_0 X_{2i}D_{f1k}^*+
     \kappa_{f} X_{bi} D_{f2k}^*)} {( c_0 X_{1i} D_{f2k}
     -\kappa_{f} X_{bi} D_{f1k})}
\eeqn
where b=3(4) for $T_{3q}=-\frac{1}{2}(\frac{1}{2})$,
$a_0=-\r2 \tan\theta_W (Q_f-T_{3f})$, $b_0=-\r2 T_{3f}$, and 
$c_0=\r2 \tan\theta_W Q_f$.
We discuss now the phases that appear in the various terms in 
$\eta_{fik}$. The term proportional to $a_0c_0$ contains the factor  
$X_{1i}^2$$D_{f1k}^*D_{f2k}$.  It is easily seen that this term 
equals $+(-)a_0c_0\frac{1}{2} Y_{1i}(\theta,\Delta \xi/2)$
$sin\theta_f e^{-i(\xi_1-\phi_f)}$, where
the +(-) sign is for $k=1(2)$.  By doing the same analysis as for the
case of the chargino contribution we find that the combinations that
arise here are  $\theta^{'}$, $\Delta \xi/2$ and $\alpha_f-\xi_1$ from which
we can construct the three combinations:
$\xi_1+\theta_{\mu}+\chi_1+\chi_2$, $\xi_2+\theta_{\mu}+\chi_1+\chi_2$
and 
 $\alpha_f+\theta_{\mu}+\chi_1+\chi_2$.
A similar analysis for the remaining terms of Eq(20) gives exactly 
the same result.
The sum of the gluino, the chargino and the neutralino exchanges
discussed above gives the total contribution from the electric dipole
operator to the quark EDM.

 The chromoelectric dipole moment $\tilde d^C$ of the quarks is
  defined via the effective dimension five operator:
\beq
{\cal L}_I=-\frac{i}{2}\tilde d^C \bar{q} \sigma_{\mu\nu} \gamma_5 T^{a} q
 G^{\mu\nu a}.
\eeq
 Contributions to $\tilde d^C$ 
of the quarks from the gluino,
 the chargino and from the neutralino exchange are given by  
\beq
\tilde d_{q-gluino}^C=\frac{g_s\alpha_s}{4\pi} \sum_{k=1}^{2}
     {\rm Im}(\Gamma_{q}^{1k}) \frac{m_{\tilde{g}}}{M_{\tilde{q}_k}^2}
      {\rm C}(\frac{m_{\tilde{g}}^2}{M_{\tilde{q}_k}^2}),
\eeq

\beq
\tilde d_{q-chargino}^C=\frac{-g^2 g_s}{16\pi^2}\sum_{k=1}^{2}\sum_{i=1}^{2}
      {\rm Im}(\Gamma_{qik})
               \frac{\mci}{M_{\tilde{q}k}^2}
                {\rm B}(\frac{\mci^2}{M_{\tilde{q}k}^2}),
\eeq

and 
\beq
\tilde d_{q-neutralino}^C=\frac{g_s g^2}{16\pi^2}\sum_{k=1}^{2}\sum_{i=1}^{4}
{\rm Im}(\eta_{qik})
               \frac{\mxi}{M_{\tilde{q}k}^2}
                {\rm B}(\frac{\mxi^2}{M_{\tilde{q}k}^2}),
\eeq
where B(r) is defined following eq.(7) and C(r) is given by 
\beq
C(r)=\frac{1}{6(r-1)^2}(10r-26+\frac{2rlnr}{1-r}-\frac{18lnr}{1-r}),
\eeq

We note that
all of the CP violating phases are contained in the factors
${\rm Im}(\Gamma_{q}^{1k})$, ${\rm Im}(\Gamma_{qik})$, and
in ${\rm Im}(\eta_{qik})$. But these are precisely the same factors 
that appear in the gluino, the chargino and the neutralino contributions to
the electric dipole operator. 

Finally we look at the CP phases that enter in the CP violating 
purely gluonic dimension six operator. 
The gluonic dipole moment $d^G$ is defined via the  
effective  dimension six operator
\beq
{\cal L}_I=-\frac{1}{6}d^G f_{\alpha\beta\gamma}
G_{\alpha\mu\rho}G_{\beta\nu}^{\rho}G_{\gamma\lambda\sigma}
\epsilon^{\mu\nu\lambda\sigma}
\eeq
where $f_{\alpha\beta\gamma}$ are the Gell-Mann coefficients, 
 $\epsilon^{\mu\nu\lambda\sigma}$
is the totally antisymmetric tensor with $\epsilon^{0123}=+1$,
and $G_{\alpha\mu\nu}$ is the gluon field strength.
Carrying out the analysis
including all phases we get 

\beq
d^G=-3\alpha_s(\frac{g_s}{4\pi m_{\tilde g}})^3
(m_t(z_1^t-z_2^t)Im(\Gamma^{12}_t)H(z_1^t,z_2^t,z_t)
+m_b(z_1^b-z_2^b)Im(\Gamma^{12}_b)H(z_1^b,z_2^b,z_b))
\eeq
\noindent 
where
\beq
\Gamma_q^{1k}=e^{-i\xi_3}D_{q2k}D_{q1k}^*,
z^q_{\alpha}=(\frac{M_{\tilde{q}\alpha}}{m_{\tilde{g}}})^2,
z_q=(\frac{m_q}{m_{\tilde{g}}})^2\nonumber\\
\eeq
It is easily seen that the combination of phases involving $\xi_3$ are 
similar to as for the gluino exchange terms discussed earlier.
Similar expressions can arise if the quarks in the loop were from the 
other two generations and one gets the combinations
$\alpha_q$+$\theta_{\mu}$+$\chi_1$+$\chi_2$,
~$\xi_3$ +$\theta_{\mu}$+$\chi_1$+$\chi_2$.

	The contribution to the neutron EDM using the non-relativistic 
	SU(6) formula is given by $d_n=\frac{1}{3}(4d_d-d_u)$.The 
	above analysis holds at the electro-weak scale. To obtain the
	value at the hadronic scale one uses renormalization group to
	evolve it down to that scale. Thus $d_n^E=\eta^E d_n$ where
	$d_n^E$  is the value at the hadronic scale and $\eta^E$ is 
	the QCD correction factor. 
The contributions of the chromoelectric dipole operator, and  
 of the purely gluonic dimension six operator to 
the quark EDMs are obtained by  use of the naive dimensional analysis, so
that $d_n^C=\frac{e}{4\pi}\tilde d_n^C \eta^C$,  and 
$d_{n}^G$=$\frac{eM}{4\pi}$ $d^G$ $\eta^G$, where 
$d_n^C$ and $d_{n}^G$ are the contributions at the electro-weak scale,
$\eta^C$ and $\eta^G$ are the 
QCD correction factors and $M=1.19$ GeV is the chiral symmetry breaking scale.

The main results of the paper are given by Eqs.(7),(11),(15) and
(19) for the contribution to the EDMs by the electric dipole operator,
by Eqs.(22)-(24) for the contribution to the EDMs by the chromo-electric
dipole operator and by Eq.(27) for the contribution to the EDM by the
purely gluonic dim 6 operator. These formulae give the contributions to
the EDMs with the most general set of CP violating phases with no 
generational mixings.

\begin{center} \begin{tabular}{|c|c|c|c|}
\multicolumn{4}{c}{Table 1: ~CP violating phases in $d_q$ and $d_{\it l}$
 in MSSM } \\
\hline
exchange  & u quark & d quark & charged  leptons \\
\hline
$\tilde g$  & $\alpha_u+\theta_1$ & $\alpha_d+
\theta_1$   &   \\

     & $\xi_3+\theta_1$& $\xi_3+\theta_1$
         &  \\
\hline
$\chi^+$  & $\alpha_d+\theta_1$ & $\alpha_u+
\theta_1$   &   \\
     & $\xi_2+\theta_1$& $\xi_2+\theta_1$
         & $\xi_2+\theta_1$ \\
\hline
$\chi^0$  & $\alpha_u+\theta_1$ & $\alpha_d+
\theta_1$   & $\alpha_{\it l}+\theta_1$\\
  & $\xi_1+\theta_1$& 
  $\xi_1+\theta_1$
         & $\xi_1+\theta_1$ \\
   & $\xi_2+\theta_1$& 
  $\xi_2+\theta_1$
         & $\xi_2+\theta_1$ \\
\hline
dim 6 & $\alpha_k+\theta_1$ & $\alpha_k+\theta_1$ &  \\
      &  k=u,d,c,s,t,b & k=u,d,c,s,t,b &  \\
\hline 
\end{tabular}
\end{center}
 The phases that enter in the quark and in the lepton
EDM's are summarized in Table 1. 
As seen from Table 1  the  electric dipole
and the chromo-electric dipole contributions to the neutron EDMs 
depend on 5 phases which can be chosen to be $\xi_i+\theta_1$ (i=1,2,3),
and $\alpha_k+\theta_1$ (k=u,d) where $\theta_1$=$\theta_{\mu}$
+$\chi_1$+$\chi_2$. The purely gluonic dimension six operator 
contribution to the quarks depends on four additional phases: 
$\alpha_k+\theta_1$ (k=t,b,c,s), and thus the neutron EDM depends on
nine independent phases. The electron EDM depends on just three  
independent phases: $\xi_i+\theta_1$ (i=1,2), and $\alpha_e+\theta_1$.
Thus the neutron and the electron EDM together depend on 10 independent 
phases. If we include the muon and the tau EDMs then the neutron and the
lepton EDMs altogether depend on twelve phases, i.e., 
  $\xi_i+\theta_1$ (i=1,2,3), $\alpha_k+\theta_1$ (k=u,d,t,b,c,s;e,$\mu$,
  $\tau$).  
  If we retain only the dominant top-stop contribution to the
 purely gluonic dimension six operator, then the total number of phases 
 reduces from 12 to 9.

	For the case of minimal supergravity, all $\alpha_k$ 
	 evolve from the phase $\alpha_0$ of $A_0$  
	 where $A_0$ is the common value of $A_i$
	at the GUT scale. Similarly all $\xi_i$
	evolve from the phase $\xi_{\frac{1}{2}}$ of the universal 
	gaugino mass $m_{\frac{1}{2}}$ at the GUT scale.
	In this case the ten  phase combinations that appear in 
	$d_n$ and $d_e$ collapse to just
	two independent ones: $\alpha_0$+$\theta_1$, and 
	$\xi_{\frac{1}{2}}$+$\theta_1$.
	 Often one sets $\xi_{\frac{1}{2}}=0$, $\chi_1=0=\chi_2$ 
	 and chooses the  independent 
	phases in minimal supergravity to be  $\alpha_0$ and $\theta_{\mu}$.
	In that limit the results of this Letter limit to the results of
	ref.\cite{us1}. 
	
	The analysis presented here is the first complete analysis of the
	neutron and of the lepton EDMs with all allowed CP phases 
	in MSSM 
	under the restriction of no generational mixings.
	In ref.\cite{us1} consistency with large CP violating phases
	was achieved by internal cancellations to satisfy the experimental
	EMD constraints with just two CP phases. In the present analysis
	since the electron EDM depends on three independent phases and the
	neutron EDM depends on nine independent phases, the satisfaction
	of the experimental EDM  constraints with large CP violating phases
	can be achieved over a much large region. 
	The analysis provides the
	framework for the investigation of the effects of large CP 
	violating phases on low energy physics.

 This research was supported in part by NSF grant 
PHY-96020274.

\end{document}